\documentstyle[preprint,tighten,eqsecnum,aps,floats,epsfig]{revtex}

\begin{document}

\preprint{CTP\#2722 \hfill April, 1998}
\draft

\title{ Relativistic y-scaling and the Coulomb sum rule in nuclei
}
\author{
M. B. Barbaro$^{(a)}$, R. Cenni$^{(b)}$,
A. De Pace$^{(a)}$, T. W. Donnelly$^{(c)}$ and A. Molinari$^{(a)}$
}
\address{
 $(a)$ Dipartimento di Fisica Teorica, Universit\`a di Torino and \\
 Istituto Nazionale di Fisica Nucleare, Sezione di Torino, \\
 via P.Giuria, I-10125 Torino, Italy\\
$(b)$ Dipartimento di Fisica, Universit\`a di Genova and \\
 Istituto Nazionale di Fisica Nucleare, Sezione di Genova, \\
 via Dodecaneso, I-16146 Genova, Italy\\
$(c)$ Center for Theoretical Physics, Laboratory for Nuclear Science and
 Department of Physics, Massachusetts Institute of Technology, \\
 Cambridge, Massachusetts 02139    
 }

\maketitle

\begin{abstract}

In this paper dividing factors $G_L$ and $G_T$ are constructed for the 
longitudinal and transverse responses of the relativistic Fermi gas in 
such a way that the reduced responses so obtained scale. These factors 
parallel another dividing factor studied previously, $H_L$, that yields a 
(different) reduced response which fulfills the Coulomb sum rule. $G_L$,
$G_T$ and $H_L$ are all found to be only very weakly model-dependent, thus
providing essentially universal dividing factors. To explore the residual
degree of dependence which remains, the scaling and sum rule properties of 
several specific models have been considered. It is seen that the 
relativistic Fermi gas (by construction) and also typical shell-model reduced 
responses successfully scale and satisfy the Coulomb sum rule, as do
experimental results at medium to high momentum transfers. On the other hand, it
is observed that the quantum hadrodynamic model does so only if 
interaction effects become weaker with increasing momentum transfer, as
predicted in the most recent versions of that model.

\end{abstract}

\vspace{1.5in}


\section{ Introduction: Reduced response functions }
\label{sec:red}

Motivated by our recent work \cite{Madre,Amore} on the Coulomb sum rule (CSR), 
in this paper we return to the related problem of $y$-scaling. In 
Refs~\cite{Madre} we devised a factor $H_L$ that, upon being divided into the 
longitudinal response function for quasielastic electron scattering,
$R_L$, produced a so-called reduced response,
\begin{equation}
r_L\equiv R_L/H_L ,
\label{eq:intro1}
\end{equation} 
with convenient energy-weighted moments.  Our approach started from
the relativistic Fermi gas (RFG) model in which the moments are well
understood. Specifically, the zeroth energy-weighted moment is the 
familiar CSR and becomes unity, that is, by construction is saturated exactly 
in the non-Pauli-blocked region where the momentum transfer $q>2 k_F$, with
$k_F$ the Fermi momentum. While the derivation of $H_L$ provided in our
previous studies was initially undertaken within the context of the RFG,
we also showed that this dividing factor is only very weakly 
model-dependent, i.e., is essentially universal.

A first motivation in the present work is to re-cast the ideas involved in
studying the behavior of the quasielastic response in the region where
the energy transfer $\omega$ is lower than its value at the quasielastic
peak, namely, in the so-called scaling region. Below we start by summarizing 
the basic ideas behind the concept of $y$-scaling \cite{Sick,Pace} in which 
one attempts to find some function of $q$ and $\omega$, here denoted $G$, 
which, when divided into the inclusive electron scattering
cross section, yields yet another reduced response with special properties.
Namely, for an appropriately chosen scaling variable $y$ (a
well-defined function of $q$ and $\omega$; see below), this reduced
response is a function only of $q$ and $y$, denoted $F(q,y)$, and
scales. The latter means that for sufficiently large momentum
transfers the function becomes universal, namely a function only of
the scaling variable $y$:
\begin{equation}
F(q,y) \stackrel{q \to \infty}\longrightarrow F(y)\equiv F(\infty,y) .
\end{equation}

The arguments for choosing the dividing function and scaling variable
may be presented from various points of view, always with the goal of
removing the single-nucleon content from the nuclear responses in as
model-independent a manner as possible while still retaining essential
relativistic effects whenever feasible. In Sec.~IIA we review the usual
approach \cite{Sick} based on the Plane-Wave Impulse Approximation (PWIA).
This yields the standard form of $y$ and $F(q,y)$. Following this review, 
in Sec.~IIB we re-cast our previous treatments of scaling \cite{Alb88} in
terms of a related dimensionless variable $\psi$ which arises naturally
when studying the RFG model. In the present work we show how $\psi$ is
directly connected to a dimensionful variable $y_{\text{RFG}}$, which for 
the RFG model is the analog of the usual $y$-variable. Thus, scaling 
behavior can be examined in terms of $y$, $y_{\text{RFG}}$ or $\psi$.
Each approach introduces specific functions by which the inclusive cross
section or the individual longitudinal and transverse response functions,
$R_L$ and $R_T$, are to be divided to yield reduced responses $F$, $F_L$ 
and $F_T$, respectively. In analogy with the previous treatment of the
CSR, these dividing functions are denoted $G$, $G_L$ and $G_T$, respectively;
that is, for example
\begin{eqnarray}
F_L &\equiv& R_L/G_L \\
F_T &\equiv& R_T/G_T .
\label{eq:intro2}
\end{eqnarray}
In presenting results in Sec.~III we show that, as was found to be the case
for the dividing factor $H_L$, the dividing factors obtained in the scaling
analysis are also only very weakly model-dependent. Indeed, in recent work
medium- and high-energy data have been tested with both the usual $y$-scaling
approach \cite{Sick} and also the RFG-motivated approach \cite{Will} 
and seen to scale successfully (see also \cite{TWDIS} for an analysis of
all existing data). Additionally, it now appears that the experimental CSR 
is reasonably well saturated at high momentum transfers \cite{Jou96}.

The second motivation for the present work is to put several models to the
test, examining both their CSR behavior and their ability to scale. Clearly
the RFG model has both properties, by construction. We also examine two
other models of the quasielastic response (see Sec.~IIC), the hybrid
model (HM) introduced in Ref.~\cite{Madre} and the quantum hadrodynamic model
(QHD) discussed in Refs.~\cite{Wal,Horo}, to test how well they satisfy the CSR
and are able to yield scaling behavior at high $q$. Each contains specific
types of interaction effects that go beyond the strict RFG model and thus
some violation of scaling and inability to yield a CSR of unity are to be
expected. Our goal is to quantify the size of such effects and to explore
whether or not the (successful) experimental behaviors can be used to 
constrain the models (see Sec.~III).

Finally, in Sec.~IV we return to summarize our findings from this study.

\section{ Formalism }
\label{sec:form}

\subsection{ The reduced response function $F$ and scaling}
\label{sec:scal-gen}

We begin the discussion of the formalism with the
most familiar chain of logic in which scaling is motivated within the
context of the PWIA for $(e,e'N)$ reactions 
\cite{Sick,Pace}. Since in this approach one uses the integrals over 
$(e,e'p)$ and $(e,e'n)$ cross sections to approximate the inclusive $(e,e')$
cross section, one may begin by performing the average of the
electron-nucleon cross section over the azimuthal angle of the ejected
nucleon: 
\begin{equation}
  {\bar\sigma}_{eN}(q,\omega;p,{\cal E}) = 
    \frac{1}{2\pi} \int d\phi_{\bbox{N}}
    \sigma_{eN} (q,\omega;p,{\cal E};\phi_{\bbox{N}}),
\label{eq:sigmabar}
\end{equation}
where $p=|\bbox{p}|$ is the missing momentum and ${\cal E}$ the missing 
energy up to an offset of the (constant) separation energy 
$E_S = m_N + M^0_{A-1} - M_A^0$. Here $m_N$ is the nucleon mass, $M^0_A$
is the target mass and $M^0_{A-1}$ is the mass of the daughter nucleon
when in its ground state (see Ref. \cite{Madre} for
the notation adopted in this work). Traditionally one uses some 
off-shell prescription for the electron-nucleon cross section, e.g., the
``cc'' prescriptions of De Forest \cite{offshell}.

Next one defines the scaling variable. The exact kinematics (i.e., no
PWIA modeling is involved --- only the imposition of energy-momentum
conservation on the $(e,e'N)$ process is required) permit one to say
that the smallest value of the missing momentum attained in the
so-called $y$-scaling region, the low-$\omega$ side of the
quasielastic peak, occurs at $p_{min}\equiv -y$, where \cite{Sick}
\begin{eqnarray}
  y &=& \frac{1}{2W^2} \left\{\left(M^0_A + \omega\right) 
    \sqrt{W^2-\left(M^0_{A-1} + m_N\right)^2} 
    \sqrt{W^2-\left(M^0_{A-1}-m_N\right)^2}\right. \nonumber \\
    && -\left.q\left[W^2+\left(M^0_{A-1}\right)^2 - m_N^2\right]\right\}
\label{eq:y-var}
\end{eqnarray}
with 
\begin{equation}
W= \sqrt{\left(M^0_A + \omega\right)^2 - q^2} .
\label{eq:defW}
\end{equation}
The energy transfer is, of course, then given as a function of $q$ and $y$.
In particular, it must lie in the range \cite{Madre} 
$\omega_t \leq \omega \leq q $, where
\begin{equation}
  \omega_t = E_S + \sqrt{(M^0_{A-1}+m_N)^2+q^2}-(M^0_{A-1}+m_N)
\end{equation}
is the threshold energy. The scaling variable vanishes when
\begin{equation}
\omega = \omega_0 = E_S + \sqrt{m_N^2 + q^2} - m_N ,
\end{equation}
which is roughly the position of the quasielastic peak, and hence the 
scaling region is characterized by having $y$ negative. It should be
stressed that no approximation is involved in using $(q,y)$ as the 
inclusive scattering variables rather than $(q,\omega)$, as they are
related in a well-defined way through Eqs.~(\ref{eq:y-var},\ref{eq:defW}). 
In discussions
of quasielastic scattering the former usually proves to be more convenient,
since the peak is found near $y=0$, providing the best reference point
for energy transfer at constant momentum transfer.

The next step usually involves making two approximations. In the first,
one assumes that protons and neutrons are distributed the same way in the
nucleus, presumably a reasonable approximation for the $N=Z$ nuclei
considered in this work. Consequently one is assuming at this point that
the spectral function of the nucleus does not carry any isospin index and
that all such dependence can be incorporated in the overall $eN$ cross
section,
\begin{equation}
  {\tilde\sigma}_{eN}(q,\omega;p,{\cal E}) \equiv \frac{E_N}{q} \left\{
      Z {\bar\sigma}_{ep}(q,\omega;p,{\cal E}) +
      N {\bar\sigma}_{en}(q,\omega;p,{\cal E}) \right\} ,
\end{equation}
where the kinematic factor $E_N/q$ with $E_N=((\bbox{q} +
\bbox{p})^2+m_N^2)^{1/2}$ has been included, as in Ref. \cite{Sick}.

Following this one attempts to remove this effective $eN$ cross section 
from under the integrals over missing energy and missing momentum involved in
going from coincidence to inclusive scattering. The usual argument is
to assume that the most important contributions to the nuclear spectral
function arise for the lowest values of $(p,{\cal E})$ that can be
reached for given values of $q$ and $y$; in the scaling region these are
${\cal E}=0$ and $p=-y$.
Thus, in this approximation, the function one hopes will scale as a 
function of $y$ when $q\rightarrow\infty$ is given by
\begin{equation}
  F(q,y) \equiv \frac{ d^2\sigma / d\Omega_e d\omega }
    { {\tilde\sigma}_{eN}(q,y;p=-y,{\cal E}=0) } .
\end{equation}
In the PWIA where the coincidence cross section factorizes into the
electron-nucleon cross section and the nuclear spectral function $\tilde S$ 
this quantity becomes
\begin{equation}
  F(q,y) = 2\pi \int\limits^Y_{-y} p \, dp\, \tilde{n} (q, y; p) ,
\label{eq:Fqy}
\end {equation}
where
\begin{equation}
  \tilde{n} (q, y; p) = \int\limits^{{\cal E}^-}_0 d{\cal E} 
    \tilde{S}(p,{\cal E}) .
\label{eq:ntilde}
\end{equation}
Here the conventional definition of the spectral function's normalization
is taken to be
\begin{equation}
\int d{\bbox{p}} \int_0^\infty d{\cal E} \tilde{S}(p,{\cal E}) = 1 .
\label{eq:norm-S}
\end{equation}
For the definition of the upper limits of integration in Eqs.~(\ref{eq:Fqy}) 
and (\ref{eq:ntilde}) we refer 
the reader to Refs.~\cite{Sick} and \cite{Madre}. Here we quote only the
asymptotic results:
\begin{equation}
  \lim_{q\to\infty} Y (q, \omega) = \infty
\end{equation}
and
\begin{equation}
  \lim_{q\to\infty} {\cal E}^- (q, y; p) = y + p -
    \left( \sqrt{M^{0^2}_{A-1} + p^2} - \sqrt{M^{0^2}_{A-1} + y^2}\right) .
\label{eq:limEminus}
\end {equation}
One sees that in the limit as $q\rightarrow\infty$ the result in 
Eq.~(\ref{eq:limEminus}) becomes independent of $q$ and hence that the
expression in Eq.~(\ref{eq:Fqy}) becomes a function only of $y$, namely, 
it scales.

\subsection{ The reduced responses $F_L^{RFG}$ and $r_L^{RFG}$}
\label{sec:scal-RFG}

Turning now to the RFG model, we have for its spectral function \cite{Madre}
\begin{equation}
  \tilde S^{\text{RFG}} (p, {\cal E}) = \frac{3 A}{8\pi k^3_F}
    \theta (k_F - p) \delta \left [{\cal E} (p) 
    - {\cal E}^{\text{RFG}} (p) \right] ,
\label{eq:support}
\end{equation}
where
\begin{equation}
{\cal E}^{\text{RFG}} (p) = \left( \sqrt{k^2_F + m_N^2}
    - \sqrt{p^2 + m_N^2} \right)
\label{eq:EFG}
\end{equation}
and where the usual convention is followed here of normalizing this spectral
function to $Z=N=A/2$:
\begin{equation}
\int d{\bbox{p}} \int_0^\infty d{\cal E} 
   {\tilde S^{\text{RFG}}}(p,{\cal E}) = A/2
\label{eq:norm-SRFG}
\end{equation}
(note the different normalization from the one assumed above). 
For the physical responses discussed below, at the end of any 
calculation we, of course, take 
$A\rightarrow\infty$ in the strict RFG model. Defining the RFG
scaling variable through the intercept of the support of the RFG spectral 
function given in Eq.~(\ref{eq:support}) and the kinematical
boundaries in the missing energy-missing momentum plane (see Ref. 
\cite{Madre}) we obtain
\begin{equation}
  y_{\text{RFG}} = m_N \zeta = m_N \left(\lambda \sqrt{1+\frac{1}{\tau}} - 
    \kappa\right) ,
\label{eq:yRFG}
\end{equation}
where, as in our past work, the dimensionless variables 
$\lambda = \omega/2m_N$, $\kappa = q/2m_N$ 
and $\tau = \kappa^2 - \lambda^2$ have been introduced. The RFG scaling 
variable in Eq.~(\ref{eq:yRFG}) vanishes for
\begin {equation}
  \lambda = \lambda_0 = \frac{1}{2} \left[ \sqrt{1 + 4\kappa^2} - 1\right]\ ,
\label{eq:lambda}
\end{equation}
namely at $(\omega_0-E_S)/2 m_N$, the value previously obtained shifted by
the separation energy (usually a small effect, and one that can be 
incorporated as in our past work \cite{Madre}). 
The different choice of scaling variable made here is motivated through
the observation that, as we have chosen to describe heavy nuclei rather
than few-body systems, we expect the main
strength in the nuclear spectral function to be found not at the lowest
values of $p$ and ${\cal E}$ that are accessible, as is indeed 
the case for few-body nuclei, but rather for values of $(p,{\cal E})$
determined (largely) by the dynamics of nucleons in other than
the $1s$-shell. 

Let us next close a logical loop and connect the RFG scaling variable
defined above to the one originally proposed for the RFG \cite{Alb88} and
referred to as $\psi$. The latter, 
in the non-Pauli-blocked region, has the following equivalent forms:
\begin{eqnarray}
  \psi &=& \frac{1}{\sqrt{\xi_F}} \left[2\theta (\lambda-\lambda_0) - 1
    \right] \left\{ \sqrt{(1 + \lambda)^2 + \frac{1}{\tau} (\tau - \lambda)^2}
    - (1 + \lambda)\right\}^{\frac{1}{2}} \nonumber \\
  &=& \frac{1}{\sqrt{\xi_F}} \left[2\theta (\lambda-\lambda_0) - 1\right]
\left\{ \kappa\sqrt{1 + \frac{1}{\tau}} - (1 + \lambda)\right\}^{\frac{1}{2}}
   \nonumber \\
  &=& \frac{1}{\sqrt{\xi_F}} \frac{\lambda - \tau}
   {\sqrt{(1+\lambda)\tau + \kappa \sqrt{\tau (1+\tau)}}} ,
\label{eq:psi}
\end {eqnarray}
where $\xi_F = \epsilon_F - 1 = \sqrt{1 + \eta_F^2} - 1$ and $\eta_F =
k_F/m_N$ are the dimensionless Fermi kinetic energy and momentum, respectively. 
Both $\psi$ and $\zeta$ vanish for $\lambda = \lambda_0$ and, by construction,
for the RFG $\psi$ is confined to lie in the range $- 1 \leq \psi \leq 1$.
It is straightforward to see that
\begin{equation}
\psi = \frac{\zeta}{\sqrt{\xi_F (1+\sqrt{1+\zeta^2})}}
\end{equation}
which implies that
\begin{equation}
 \xi_F \psi^2 = \sqrt{1 + \zeta^2} - 1 .
\label{eq:psi2}
\end{equation}
The physical significance of $\psi$ is then immediately
apparent: among the nucleons responding to an external
probe one has the smallest kinetic energy and this is given
by $\psi^2$ (in units of the dimensionless Fermi kinetic energy $\xi_F$).

We may now proceed as in the previous subsection and divide by a factor
that is proportional to the sum of the $ep$ and $en$ cross sections, 
weighted by $Z$ and $N$, respectively. From our previous work on the RFG
model \cite{Alb88} we know that the azimuthal-angle-averaged $eN$ cross section 
(compare Eq.~(\ref{eq:sigmabar})) is proportional to $v_L f_L + v_T f_T$,
where $v_L$ and $v_T$ are the usual longitudinal and transverse Rosenbluth
kinematical factors and, in terms of the electric $G_{Ep,n}$ and 
magnetic $G_{Mp,n}$ Sach's form factors, we have
\begin{eqnarray}
f_L (\kappa,\lambda;\chi) &=& \frac{\kappa^2}{\tau} \left[ Z ( G_{Ep}^2
   + W_{2p} \chi^2 ) + N ( G_{En}^2 + W_{2n} \chi^2 ) \right] \\
f_T (\kappa,\lambda;\chi) &=& Z ( 2\tau G_{Mp}^2 + W_{2p} \chi^2)
   + N ( 2\tau G_{Mn}^2 + W_{2n} \chi^2)
\label{eq:littlef}
\end{eqnarray}
with
\begin {equation}
  W_{2p,n}(\tau) = \frac{1}{1 + \tau} \left[G_{Ep,n}^2 (\tau) 
   + \tau G^2_{Mp,n} (\tau)\right] .
\label{eq:WRFG}
\end{equation}
Here $\chi\equiv\eta\sin\theta$, where $\eta\equiv p/m_N$ and 
$\theta$ is the angle between {\bf p} and {\bf q}. Note that
the $\delta$-function in Eq.~(\ref{eq:support}) requires that
${\cal E}={\cal E}^{\text{RFG}}$ as defined in Eq.~(\ref{eq:EFG}).

We could proceed to follow exactly the procedures outlined in the
previous subsection, working from the unseparated inclusive cross 
section towards a reduced response which, if successful, would scale
as $q \to \infty$. Instead, in this work for the most part we work directly 
with the separated longitudinal and transverse responses, $R_L$ and $R_T$,
respectively, since (1) we are most interested in
model-to-model comparisons and the same procedures may be followed
in each case (i.e., focusing on $L$ or $T$ responses directly), (2) a
few cases exist where $L/T$ separations have been performed experimentally,
and (3) we wish to draw comparisons with
studies of the Coulomb sum rule where only the $L$ response is relevant.
In fact, in most of the discussions to follow we shall limit our attention
to the longitudinal channel and only give a few results for the transverse
case.

We now seek reduced responses denoted $F_{L,T} (\kappa,\psi)$
that scale. These are to be obtained from the inclusive response functions
$R_{L,T} (\kappa,\lambda)$ by dividing through by specific functions
denoted $G_{L,T} (\kappa,\lambda)$ (discussed below):
\begin{equation}
F_{L,T} (\kappa,\psi)\equiv R_{L,T} (\kappa, \lambda)
  /G_{L,T} (\kappa, \lambda) .
\label{eq:defG}
\end{equation}
If the dividing functions are chosen appropriately, then as above the 
reduced responses
defined in Eq.~(\ref{eq:defG}) will scale, namely, become functions only of
a single scaling variable such as $\psi$ defined above when 
$\kappa\rightarrow\infty$,
\begin{equation}
F_{L,T} (\kappa,\psi) \stackrel{\kappa \to \infty}\longrightarrow F_{L,T} 
(\psi)\equiv F_{L,T} (\infty,\psi) .
\end{equation}

To obtain the dividing functions we proceed from the spectral function for
the RFG, using the general PWIA expressions given above. Let us first focus
on the longitudinal response. With Eq.~(\ref{eq:support}) we have
\begin{equation}
  \tilde n^{\text{RFG}} (q, y_{\text{RFG}}, p) = \int\limits_0^{{\cal E}^-}
    d{\cal E} \tilde S^{\text{RFG}} (p,{\cal E}) = \frac{3 A}{8 \pi k^3_F}
    \theta (k_F - p) \theta (p + y_{\text{RFG}})
\end{equation} 
and therefore, from Eq.~(\ref{eq:Fqy}), together with a factor $Z=N=A/2$ to
account for the different normalizations of $\tilde S$ and 
$\tilde S^{\text{RFG}}$, we have
\begin{equation}
  F_L^{\text{RFG}} (q, y_{\text{RFG}}) \equiv F_L^{\text{RFG}}(y_{\text{RFG}}) =
    \frac{3}{2 k^3_F} \int\limits_{-y_{\text{RFG}}}^{k_F}
    p\,dp = \frac{3}{4 k^3_F} (k^2_F - y^2_{\text{RFG}}) .
\end{equation}
By exploiting Eq.~(\ref{eq:psi2}) it is then an easy matter to obtain
\begin{equation}
  k^2_F - y^2_{\text{RFG}} = m_N^2 \xi_F(1 - \psi^2) [2 + \xi_F (1 + \psi^2)]
\end{equation}
and hence the scaling function of the RFG will read
\begin{equation}
  F_L^{\text{RFG}}(\psi) = \frac{3\xi_F}{2m_N \eta^3_F} 
    (1 - \psi^2)\theta(1 - \psi^2)\left[1+\frac{1}{2} \xi_F(1+\psi^2)\right] .
\label{eq:RFGscale}
\end {equation}

We now recall from previous work \cite{Alb88} that the RFG longitudinal
response function is given by (see Fig.~\ref{fig:Fig1})
\begin{equation}
  R_L^{\text{RFG}} (\kappa, \lambda) = \frac{3\xi_F}{4m_N \kappa \eta^3_F} 
    [Z U_{Lp}+N U_{Ln}]
    (1 - \psi^2) \theta (1 - \psi^2) ,
\label{eq:RLRFG}
\end{equation}
where the nucleonic terms $U_{Lp,n}$ are the following:
\begin{equation}
  U_{Lp,n} = \frac{\kappa^2}{\tau} \left[G_{Ep,n}^2 (\tau) 
   + W_{2p,n}(\tau) \Delta\right] .
\label{eq:URFG}
\end{equation}
Furthermore $\Delta$, which, in the non-Pauli-blocked domain, reads
\begin{equation}
  \Delta = \frac{\tau}{\kappa^2} \left[ \frac{1}{3} \left( \epsilon^2_F 
    + \epsilon_F \sqrt{1 + \zeta^2} + 1 + \zeta^2\right) 
    + \lambda \left(\epsilon_F + \sqrt{1 + \zeta^2}\right) + \lambda^2\right]
    - (1 + \tau) ,
\label{eq:Delta}
\end{equation}
represents the transverse component (with respect to $\bbox{q}$) of the
momentum of the struck nucleon \cite{Amore}. 
It follows that the required RFG dividing function is given by
\begin{eqnarray}
G_L (\kappa,\lambda) &=& \frac{Z U_{Lp}+N U_{Ln}}
    {2\kappa [1+\xi_F (1+\psi^2)/2]} \\
  &=& \frac{1}{2\kappa} (Z U_{Lp}+N U_{Ln}) + {\cal O}(\xi_F) ,
\label{eq:RFG-G}
\end{eqnarray}
telling us that the scaling function of the RFG arises from $R_L$ not only 
by pulling out the non-scaling single-nucleon factor involving 
$U_{Lp}/\kappa$ and $U_{Ln}/\kappa$, but also contains a small correction
for the medium dependence in the problem (the factor in the denominator
involving $\psi$). Typically $\xi_F$ is very small, attaining values
as large as 0.04 only in the heaviest of nuclei, and thus both the
correction in the denominator containing $\xi_F$ and the medium-dependent
effects in the numerator embodied in $\Delta$ provide only few percent
corrections at high momentum transfers, as long as $\psi$ is not permitted 
to become large. Indeed, in the RFG its magnitude is bounded by unity
and so this always obtains at high $q$; if the ideas here are carried
over to other models where large excursions away from the quasielastic
peak are permitted, then one should check the actual size
of these medium-dependent corrections.

Dividing the longitudinal response function by $G_L$ then
yields the reduced response $F_L$ which, at least for the RFG, scales with 
$\psi$, $\zeta$ or $y_{\text{RFG}}$. Note that we can use 
Eq.~(\ref{eq:psi2}) to write
\begin{equation}
  1 + \frac{1}{2} \xi_F (1 + \psi^2) = \frac{1}{2} \left(\epsilon_F + 
    \sqrt{1 + \zeta^2}\right) ,
\label{eq:denpsizeta}
\end {equation}
giving us some insight into 
the physical significance of the denominator in Eq.~(\ref{eq:RFG-G}): namely, 
it represents an average energy of the nucleons responding to the external 
probe. 

Proceeding in exactly the same way for the transverse channel it is
straightforward to see that the dividing factor required in discussions
of scaling in that case is
\begin{eqnarray}
G_T (\kappa,\lambda) &=& \frac{Z U_{Tp}+N U_{Tn}}
    {2\kappa [1+\xi_F (1+\psi^2)/2]} \\
 &=& \frac{1}{2\kappa} (Z U_{Tp}+N U_{Tn}) + {\cal O}(\xi_F) ,
\label{eq:RFG-GT}
\end{eqnarray}
where (compare Eq.~(\ref{eq:littlef}))
\begin{equation}
  U_{Tp,n} = 2\tau G_{Mp,n}^2 (\tau) 
   + W_{2p,n}(\tau) \Delta .
\label{eq:URFGT}
\end{equation}
The unseparated responses may then be analyzed using
\begin{equation}
F (\kappa, \psi) \equiv \frac{d^2 \sigma/d \Omega_e d \omega}
   {\sigma_M [v_L G_L (\kappa, \lambda) + v_T G_T (\kappa, \lambda)]} ,
\label{eq:totscal}
\end{equation}
where as usual one would have for the cross section
\begin{equation}
d^2 \sigma/d \Omega_e d \omega = \sigma_M [v_L R_L (\kappa, \lambda) 
   + v_T R_T (\kappa, \lambda)]
\end{equation}
with $\sigma_M$ the Mott cross section and $v_{L,T}$ the usual
Rosenbluth kinematical factors. 
Naturally, in the RFG model we have $F^{\text{RFG}}=F^{\text{RFG}}_T=
F^{\text{RFG}}_L$, the result being given in Eq.~(\ref{eq:RFGscale}).

The developments presented here for scaling have been motivated by our 
previous discussions of the Coulomb sum rule and various energy-weighted
moments of another reduced response denoted $r_L (\kappa, \lambda)$
\cite{Madre,Alb88}. In those studies the longitudinal response $R_L (\kappa,
\lambda)$ was divided by a function $H_L (\kappa, \lambda)$ to yield
\begin{equation}
r_L (\kappa,\psi)\equiv R_L (\kappa, \lambda)/H_L (\kappa, \lambda) ,
\label{eq:defH}
\end{equation}
where the n$^{th}$ moment of the longitudinal response of the nucleus is
given by
\begin{equation}
  \Xi^{(n)} = \int\limits_0^\kappa d\lambda \, \lambda^n \, 
   r_L (\kappa, \lambda) .
\end{equation}
In particular, the $n=0$ moment, $\Xi^{(0)}$, is the Coulomb sum 
rule. In the case of the RFG our previous work showed that
\begin{equation}
  H_L (\kappa, \lambda) = \frac{Z U_{Lp}+N U_{Ln}}{J_L} ,
\label{eq:HL}
\end{equation}
where $U_{Lp,n}$ are given above and where
\begin{eqnarray}
  J_L &=& \frac{\kappa\eta_F^3}{2\xi_F} \frac{\partial\psi}{\partial\lambda}
  \\
 &=& \left( \frac{\kappa^2}{\tau} \right) 
    \left( \frac{1+2\lambda}{1+\lambda} \right) + {\cal O}(\xi_F) .
\end{eqnarray}
Thus, by dividing the charge response of the RFG by Eq.~(\ref{eq:HL}) 
(see Eq.~(\ref{eq:RLRFG}))
one obtains the following reduced longitudinal response
\begin{equation}
  r^{\text{RFG}}_L (\kappa, \lambda) = \frac{3}{8 m_N } 
    (1 - \psi^2) \theta 
    (1 - \psi^2) \frac{{\partial\psi}}{\partial\lambda} ,
\end{equation}
which, by construction, fulfills the CSR in the non-Pauli-blocked domain,
as can easily be verified. Observe also that while $F^{\text{RFG}}_L$ 
attains its maximum for $\psi = 0$,
i.~e. for $\lambda = \lambda_0$, this is nearly, but not exactly, true for
$r^{\text{RFG}}_L$ only for small values of $\eta_F$. Indeed using the simple 
expression
\begin{equation}
  \psi \cong \frac{1}{\eta_F} \left[ \frac{\lambda(\lambda+1)}{\kappa} 
   - \kappa\right] ,
\label{eq:psiappr}
\end{equation}
which approximates Eq.~(\ref{eq:psi}) quite faithfully for very large values 
of $\kappa$ except on the borders of the response region, it turns out that the 
maximum of $r^{\text{RFG}}_L$ occurs at
\begin {equation}
  \lambda = \frac{1}{2} \left[ \sqrt{1 + 4 \kappa^2\left(1+\frac{1}{\sqrt 5} 
    \frac{\eta_F}{\kappa}\right)} - 1\right] = \lambda_0 + {\cal O}(\eta_F^2) .
\end {equation}

Before leaving this discussion of the RFG model reduced responses, for
completeness let us
mention another version of the reduced longitudinal response that can be 
obtained in terms of the variable $\zeta$ rather than $\psi$ as in 
(\ref{eq:defH}). For this purpose one has to divide 
Eq.~(\ref{eq:RLRFG}) by Eq.~(\ref{eq:HL}) with $J_L$ replaced by
\begin{equation}
  J_L^\prime = \frac{2\kappa\eta_F^2}{3} \frac{\partial \zeta}{\partial\lambda}
    \frac{1}{\displaystyle\epsilon_F - 
    \frac{\strut\sinh^{-1}\eta_F}{\displaystyle\eta_F}} .
\label{eq:JLp}
\end{equation}
The result is
\begin{equation}
  r^{\text{RFG}'}_L(\kappa, \lambda) = \frac{1}{2m_N \eta_F} 
    \frac{\epsilon_F - \sqrt{1 + \zeta^2}}{\displaystyle\epsilon_F -
    \frac{\strut\sinh^{-1}\eta_F}{\displaystyle\eta_F}}\theta(\eta_F -|\zeta|)
    \frac{\partial\zeta}{\partial\lambda} ,
\end{equation}
which again fulfills the CSR. Indeed
\begin{equation}
  \Xi^{(0),\text{RFG}} = 
    \int\limits_0^\kappa d\lambda\, r_L^{\text{RFG}'} (\kappa, \lambda) = 
    \frac{1}{\eta_F} \frac{1}{\left(\displaystyle\epsilon_F - 
    \frac{\strut\sinh^{-1}\eta_F}{\displaystyle\eta_F}\right)}
    \int_{-\eta_F}^{\eta_F} d\zeta 
    \left(\epsilon_F - \sqrt{1+\zeta^2}\right) = 1 .
\end{equation}

\subsection{ Scaling and sum rules in the HM and QHD models}
\label{sec:scal-HMQHD}

\begin{figure}[tb]
\begin{center}
\mbox{\epsfig{file=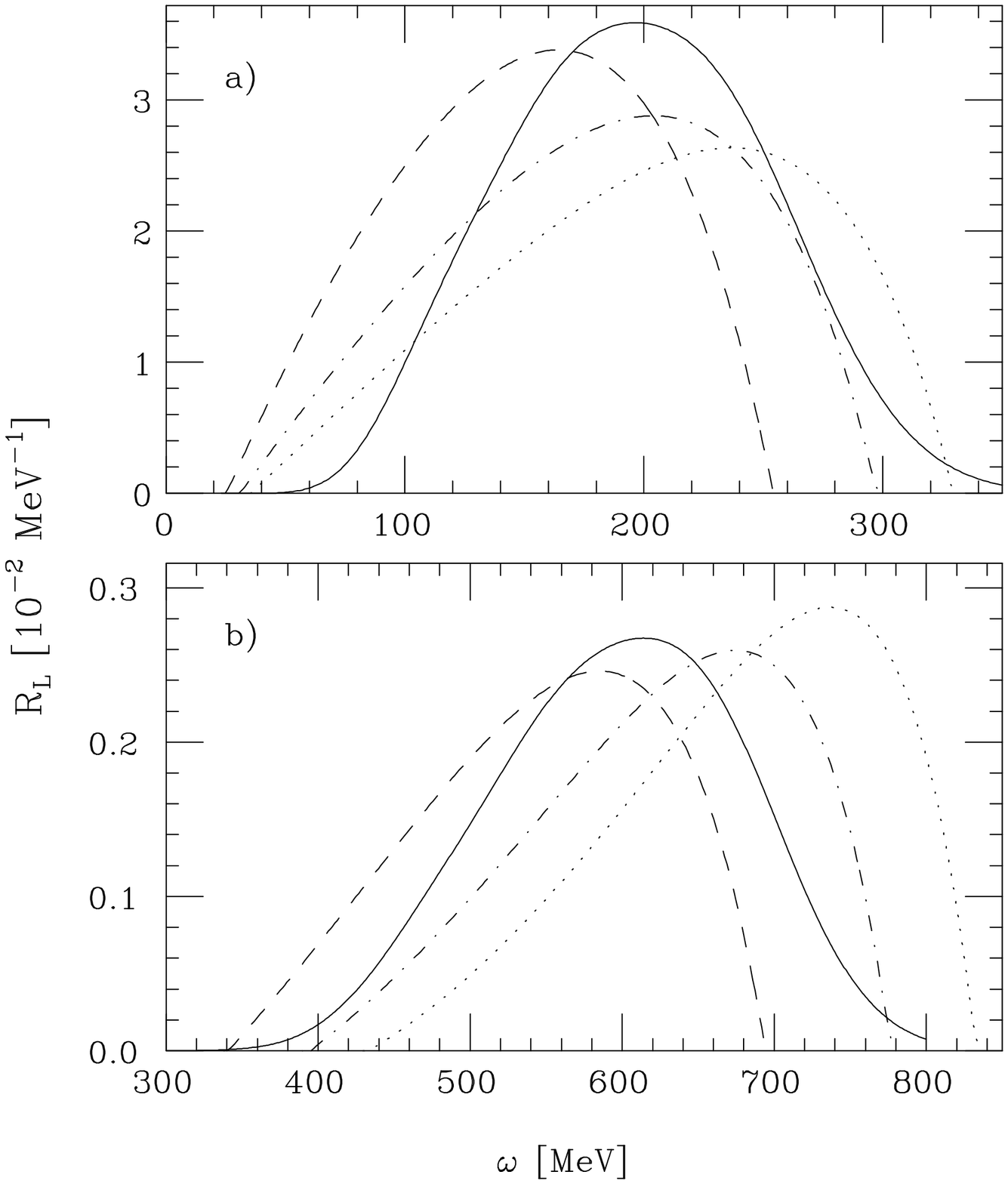,width=0.9\textwidth}}
\caption{The longitudinal response function $R_L$ is displayed versus $\omega$
for $q=0.55$ GeV/c in panel (a) and $q=1.14$ GeV/c in panel (b). The four 
curves shown are the
following: dashed --- RFG, solid --- HM, dot-dashed ---
QHD model ($m_N^\ast$=0.8 $m_N$) and dotted --- ($m_N^\ast$=0.68 $m_N$). 
Here and in all of the following figures the nucleus chosen is $^{40}$Ca. 
For the RFG and QHD models we always assume $k_F$=230 MeV/c, while the HM
parameters are discussed in the text.
}
\label{fig:Fig1}
\end{center}
\end{figure}

An extension of the RFG to account for the binding of the nucleons inside
the nucleus, thus curing a flaw of the RFG related to its negative
separation energy, is represented by the hybrid model studied previously in 
Ref.~\cite{Madre}. The HM has continuum states which are simply plane
waves, as in the RFG model, but has bound states described by shell-model
wave functions obtained by solving the Schr\"odinger equation with some
choice of potential well. For simplicity in studying the $A \to \infty$ 
limit, in \cite{Madre} we chose to consider harmonic oscillator bound-state 
wave functions. The longitudinal response $R^{\text{HM}}_L$ is
also shown in Fig.~\ref{fig:Fig1}. 

In the HM scaling variables 
which incorporate the shift result from using Eq.~(\ref{eq:yRFG}) for
$\zeta'$ (and a corresponding dimensionful variable $y'$) or 
Eq.~(\ref{eq:psi})
for $\psi'$ by making the replacements $\lambda \to \lambda^\prime$ and 
$\tau \to \tau^\prime = \kappa^2 - {\lambda^\prime}^2$, where
\begin{equation}
  \lambda^\prime = \lambda - \lambda_{\text{shift}} ,
\label{eq:shift}
\end {equation}
being
\begin{equation}
\lambda_{\text{shift}} = \frac{1}{2m_N} (T_F + E_S) 
\label{eq:shiftlam}
\end{equation}
and $T_F=\sqrt{k_F^2 + m_N^2}-m_N = m_N\xi_F $ the Fermi kinetic energy.
The HM turns out to have the width of its reduced response
(its variance, see Ref.~\cite{Madre}) identical to a RFG
computed with a Fermi momentum somewhat larger than the usual one
(237 MeV/c for the HM, versus the 230 MeV/c value for $k_F$ used here for the
RFG to correspond  
to nuclei near $^{40}$Ca or $^{56}$Fe). In conformity, $\psi^\prime$ should be
computed with $k_F$=237 MeV/c as well. As seen in Fig.~\ref{fig:Fig1},
the shift arising from replacing $\lambda$ with $\lambda_{\text{shift}}$ 
according to Eq.~(\ref{eq:shiftlam}) is apparent.

Scaling may then be examined for the HM by computing
\begin{equation}
F^{\text{HM}}_L (\kappa,\psi)\equiv 
  R^{\text{HM}}_L (\kappa, \lambda)/G_L (\kappa, \lambda)
\label{eq:F-HM}
\end{equation}
and the various energy-weighted moments of the longitudinal response, 
including the zeroth moment or CSR, may be computed using 
\begin{equation}
r^{\text{HM}}_L (\kappa,\psi)\equiv 
  R^{\text{HM}}_L (\kappa, \lambda)/H_L (\kappa, \lambda) .
\label{eq:r-HM}
\end{equation}
Note that we have used the {\em same} dividing factors $G_L$ and $H_L$ that
were developed from our discussions of the RFG in the previous subsection.
In \cite{Madre} we showed that $H_L$ is universal in that the form found
for the HM essentially coincides with the one obtained above. In the next
section we present results where the corresponding universality of $G_L$
is tested numerically. Naturally these explorations also involve displaying
the reduced responses given in Eqs.~(\ref{eq:F-HM}) and (\ref{eq:r-HM}) as
functions of $\psi'$ as well as $\psi$.
 
Finally we turn to an examination of the QHD model \cite{Wal}. 
In this model protons and neutrons in the nucleus are described by 
Dirac spinors and move in strong Lorentz scalar
and vector mean fields. These in turn arise self-consistently from the exchange
of $\sigma$ and $\omega$ mesons between the same nucleons on which
they act. The scalar field dresses the bare mass of the nucleon, 
considerably lowering its value; the vector field uniformly shifts
the fermion spectrum. As a consequence the QHD charge response of
nuclear matter in Hartree approximation is unaffected by the vector
field, while it turns out to be quite sensitive to the effective mass
$m^\ast_N$ induced by the scalar field, as shown in Ref.~\cite{Horo}.
This is, of course, true in the simple approximation of constant relativistic
mean fields. An improved description allows for an energy dependence of the
latter which helps to account for the data of proton-nucleus elastic scattering.
We shall however ignore these refinements in the present analysis.

In Fig.~\ref{fig:Fig1}, in order to directly compare with the 
results of Ref.~\cite{Horo}, we display the longitudinal response of 
nuclear matter, again using $k_F = 230$ MeV/c, in the Hartree 
version of QHD for $q = 0.55$ and 1.14 GeV/c, respectively.
In addition to the RFG and HM results discussed above, here we also show 
curves corresponding to $m^\ast_N = 0.8\, m_N$ and $m^\ast_N = 0.68\, m_N$,
namely, those displayed in Ref.~\cite{Horo} for the case of constant 
relativistic mean fields. From the figure the sensitivity of the charge 
response to $m_N^\ast$ is fully apparent. Clearly the more $m^\ast_N$ 
deviates from $m_N$ the more the longitudinal QHD response is hardened
(shifted to higher values of $\omega$). The most recent version of QHD
suggests that as $q$ increases the effective value of $m^\ast_N$  should
tend towards $m_N$ and thus that, for intermediate-energy studies
($q\sim$ 500 MeV/c), $m^\ast_N/m_N$ may be somewhere between 0.68 and 0.8,
while for high $q$, a value between 0.8 and 1.0 is likely to be preferred.

As above for the HM model, scaling will be examined in the QHD model by 
computing
\begin{equation}
F^{\text{QHD}}_L (\kappa,\psi)\equiv 
  R^{\text{QHD}}_L (\kappa, \lambda)/G_L (\kappa, \lambda)
\label{eq:F-QHD}
\end{equation}
and likewise the various energy-weighted moments of the longitudinal response
will be computed using 
\begin{equation}
r^{\text{QHD}}_L (\kappa,\psi)\equiv 
  R^{\text{QHD}}_L (\kappa, \lambda)/H_L (\kappa, \lambda) ,
\label{eq:r-QHD}
\end{equation}
where $R^{\text{QHD}}_L (\kappa,\psi)$ is given below in 
Eq.~(\ref{eq:RLRFGast}).
Our contention is that the dividing factors are (at least to a very good level 
of approximation; see Sec.~III) universal and accordingly here we have 
continued to use the {\em same} dividing factors $G_L$ and $H_L$ that
were developed from our discussions of the RFG in the previous subsection.
In discussing the QHD results in the next section we shall display the
reduced responses in Eqs.~(\ref{eq:F-QHD}) and (\ref{eq:r-QHD}) as functions
both of $\psi$ and also $\psi^{\ast}$, namely, the RFG scaling variable
given in Eq.~(\ref{eq:psi}) with $m_N$ replaced by $m_N^\ast$. We shall also 
briefly explore what
happens when the $m_N \to m_N^{\ast}$ replacement is made in a particular
way to obtain modified dividing factors, $G_L^{\ast}$ and $H_L^{\ast}$, 
defined below.

Before proceeding to reduce the QHD response a delicate point arises
in connection with the question: does the concept of effective mass also
apply to the single-nucleon content of the longitudinal response? The 
answer is yes and no. In fact the approach of Ref.~\cite{Horo}
requires the Dirac $F_1$ and Pauli $F_2/m_N$ form factors in the
medium to be {\em identical} to those in free space (or, in other words,
the nucleon current to be unaltered). Thus they are expressed in
terms of the Sach's electric and magnetic form factors through the
standard relations
\begin{mathletters}
\begin{equation}
  F_{1p,n}(\tau) = \frac{1}{1 + \tau} \left[ G_{Ep,n}(\tau) 
   + \tau G_{Mp,n}(\tau)\right]
\end{equation}
and
\begin{equation}
  F_{2p,n}(\tau) = \frac{1}{1 + \tau} \left[ G_{Mp,n}(\tau) 
   - G_{Ep,n}(\tau)\right] ,
\end{equation}
\end{mathletters}
with $G_{Ep,n}$ and $G_{Mp,n}$ fitting the results of electron-nucleon elastic
scattering in {\em free} space. Then, to keep $F_{1p,n}$ and $F_{2p,n}/m_N$ 
unaltered, the Sach's form
factors in the medium should be given by
\begin{mathletters}
\label{eq:GEMstar}
\begin{eqnarray}
  G_{Ep,n}^\ast (q, \omega) &=& F_{1p,n}(\tau) 
   - \tau^\ast\frac{m_N^\ast}{m_N}  F_{2p,n}(\tau)
   , \\
  G_{Mp,n}^\ast (q, \omega) &=& F_{1p,n}(\tau) 
   + \frac{m_N^\ast}{m_N} F_{2p,n}(\tau) ,
\label{eq:GMstar}
\end{eqnarray}
\end{mathletters}
where $\tau^\ast = (m_N/m_N^\ast)^2 \tau$.
Hence they will differ from the corresponding quantities in free space.
Indeed it is the $m_N^\ast$ appearing in Eq.~(\ref{eq:GMstar}) that causes a 
strong reduction of the proton's convection current and, as a consequence,
leads a considerable quenching of the magnetic moment of odd-proton-nuclei.
To restore the agreement between the theoretical predictions of QHD and
experiment a substantial back-flow current contribution has then
to be invoked \cite{Shep}.

For the quasielastic charge response we face a similar situation. In fact, as 
discussed
above, we reduce the QHD longitudinal response by using as dividing factors 
$G_L$ and $H_L$ defined in Eqs.~(\ref{eq:RFG-G}) and (\ref{eq:HL}), 
respectively, with the {\em bare} nucleon mass and, hence, with the {\em free} 
$G_{Ep,n}$ and $G_{Mp,n}$. An alternative might be to use $G_{Ep,n}^\ast$
and $G_{Mp,n}^\ast$ given above, together with
\begin{equation}
  W^\ast_{2p,n} (q, \omega)  = \frac{1}{1 + \tau^\ast}
    \left(G_{Ep,n}^{\ast^2} + \tau^\ast G_{Mp,n}^{\ast^2}\right) ,
\end {equation}
to define $U_L^\ast$ using an expression that is analogous to 
Eq.~(\ref{eq:WRFG}).
This would lead to dividing factors we denote $G_L^\ast$ and $H_L^\ast$,
and, by making the replacement $m_N \to m_N^\ast$ in Eq.~(\ref{eq:psi}), to
a corresponding scaling variable $\psi^\ast$. Accordingly we would have
different reduced responses (again labeled with an asterisk):
\begin{equation}
F^{\text{QHD}\ast}_L (\kappa,\psi)\equiv 
  R^{\text{QHD}}_L (\kappa, \lambda)/G_L^\ast (\kappa, \lambda)
\label{eq:F-QHDast}
\end{equation}
and 
\begin{equation}
r^{\text{QHD}\ast}_L (\kappa,\psi)\equiv 
  R^{\text{QHD}}_L (\kappa, \lambda)/H_L^\ast (\kappa, \lambda) .
\label{eq:r-QHDast}
\end{equation}
Clearly, since the QHD longitudinal response function is given by (compare with 
Eq.~(\ref{eq:RLRFG}))
\begin{equation}
  R_L^{\text{QHD}} (\kappa, \lambda) = 
    \frac{3 \xi_F^\ast}{4m_N \kappa{\eta_F^\ast}^3} 
    [Z U^\ast_{Lp}+N U^\ast_{Ln}]
    (1 - {\psi^\ast}^2) \theta (1 - {\psi^\ast}^2) \ ,
\label{eq:RLRFGast}
\end{equation}
the reduced responses $F^{\text{QHD}\ast}_L$ and $r^{\text{QHD}\ast}_L$  will
behave as functions of $\psi^\ast$ exactly as
$F^{\text{RFG}}_L$ and $r^{\text{RFG}}_L$  behave as functions of
$\psi$. In the following section we return briefly to examine the 
consequences of these alternative definitions of the dividing factors.

\section{Results}
\label{sec:appl}

We have thus reached the point at which the longitudinal reduced response 
functions $F_L$ (for scaling discussions) and $r_L$ (for sum-rule discussions)
are to be obtained by dividing $R_L$ by two different factors, $G_L$ and 
$H_L$, respectively. The first is given by Eq.~(\ref{eq:RFG-G}), while the 
second is given in Eq.~(\ref{eq:HL}) and they are related according to
\begin {eqnarray}
  G_L (\kappa, \lambda) &=& \left( \frac{\eta^3_F}{4\xi_F} \right)
    \frac{\partial \psi}{\partial\lambda}
    \frac{1}{1 + \xi_F (1 + \psi^2)/2 } H_L (\kappa, \lambda) \\
&=& \frac{1}{2} \left( \frac{\kappa}{\tau} \right) 
    \left( \frac{1+2\lambda}{1+\lambda} \right) 
    H_L (\kappa, \lambda)  + {\cal O}(\xi_F) .
\end{eqnarray}

\begin{figure}[tb]
\begin{center}
\mbox{\epsfig{file=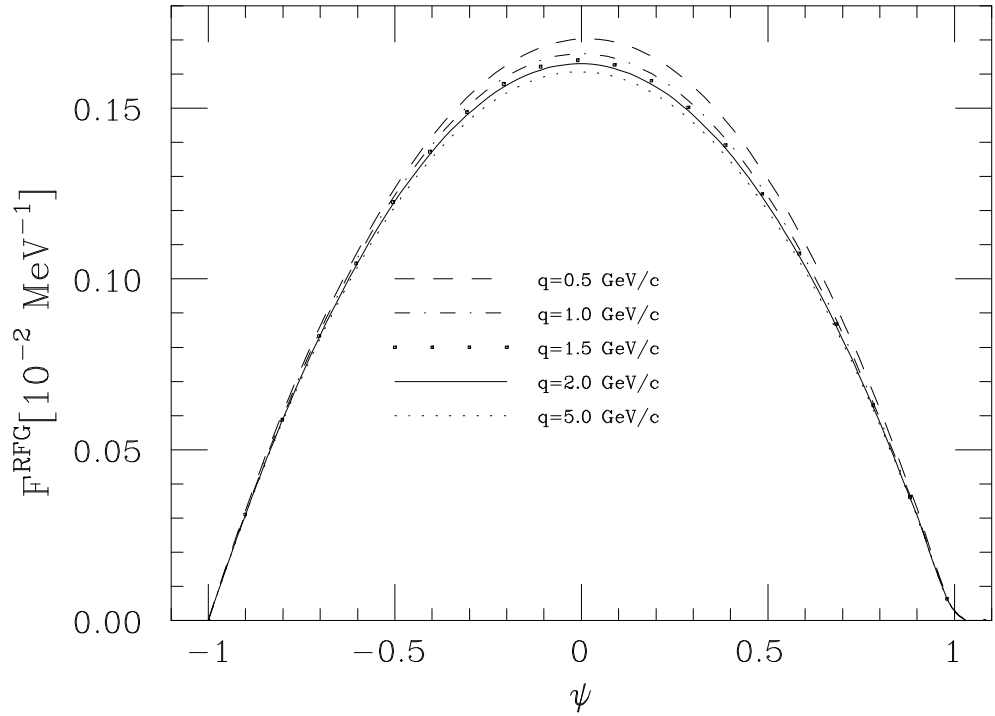,width=0.9\textwidth}}
\caption{ Scaling is shown as a function of $\psi$ using the RFG cross 
section at 10$^o$ scattering angle divided by the off-shell cc1 form 
with separation energy $E_S =8$ MeV for a wide range of momentum transfers.
}
\label{fig:Fig2}
\end{center}
\end{figure}

\begin{figure}[tb]
\begin{center}
\mbox{\epsfig{file=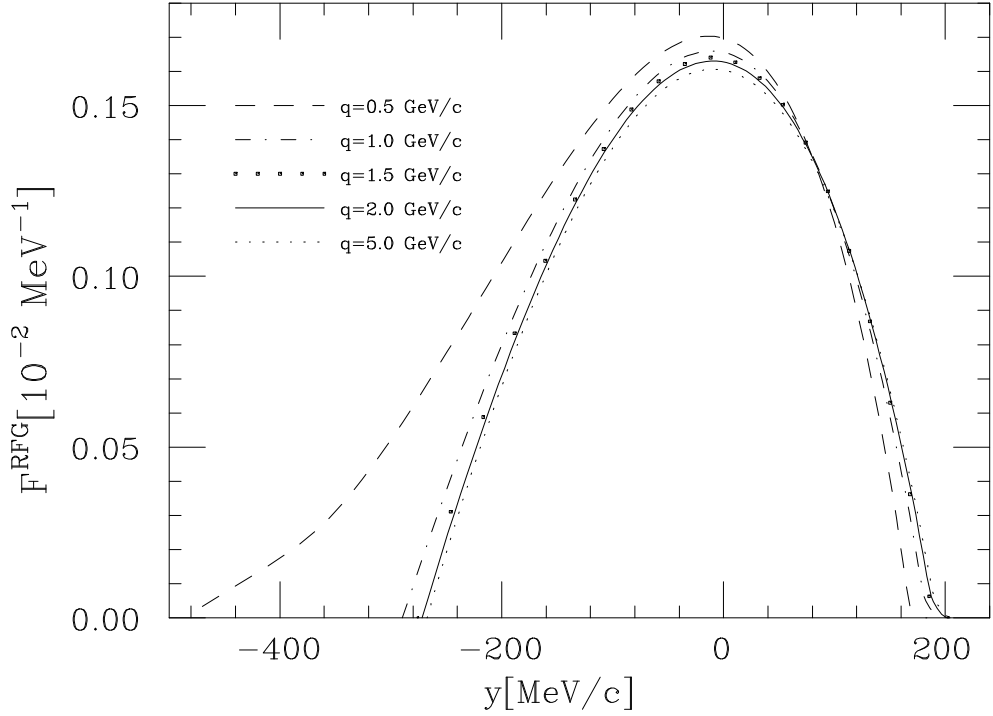,width=0.9\textwidth}}
\caption{The scaling function $F^{\text{RFG}}$ is shown for the same
conditions employed in Fig.~\ref{fig:Fig2} as a function
of the $y$-scaling variable defined in Eq.~(\ref{eq:y-var}), rather than
as a function of $\psi$ (see Eq.~(\ref{eq:psi})) in which it scales
exactly, by construction. The separation energy occurring in the definition
of $y$ has been set to 8 MeV for these results.
}
\label{fig:Fig3}
\end{center}
\end{figure}

Naturally, the analogous dividing function $G_T$ in Eq.~(\ref{eq:RFG-GT}) should
be used when treating the separated transverse response or the combination
${\sigma_M [v_L G_L (\kappa, \lambda) + v_T G_T (\kappa, \lambda)]}$ in 
Eq.~(\ref{eq:totscal}) may be used when dealing with an unseparated
cross section. To assess the universality of using the various approaches
discussed in the preceeding section, we show in Figs.~\ref{fig:Fig2} and
\ref{fig:Fig3} two mixed representations of the usual $y$-scaling results
and the RFG-motivated results. In Fig.~\ref{fig:Fig2} the total RFG cross
section is displayed divided by the De Forest cc1 off-shell result 
\cite{offshell} --- this is what is usually set as the dividing factor 
in presenting unseparated data in the form $F(q,y)$ versus $y$, as in 
Ref.~\cite{Sick}. Here we have taken $E_S=8$ MeV as being typical of
nuclei across the periodic table. Clearly once the
momentum transfer is large enough (say above about 1 GeV/c) it is 
irrelevant whether the cc1 form is used or the RFG form is employed: the
dividing factor is effectively universal. In Fig.~\ref{fig:Fig3} the usual
reduced response for the RFG (i.e., with the dividing factor given in 
Eq.~(\ref{eq:totscal})) is displayed, however as a function of $y$ rather
than $\psi$, in which it of course scales perfectly. Again, at all but the 
lowest momentum transfer the scaling is excellent in $y$ as well as $\psi$.
These and other more extensive studies show that at high momentum transfers
the choice of scaling variable is largely irrelevant, as long as one of 
the relativistic forms discussed in this work is employed, and that
the on- and off-shell prescriptions for the dividing factors yield
essentially model-independent results for scaling (as they did for the CSR).

By exploiting the essentially universal dividing factors $H_L$ and $G_L$, 
we are thus in a position to test whether or not a given model fulfills 
the CSR and scales. Here we examine the longitudinal response functions of 
the HM \cite{Madre} and QHD \cite{Wal,Horo} models in concert with the RFG 
results obtained previously.

\begin{figure}[tb]
\begin{center}
\mbox{\epsfig{file=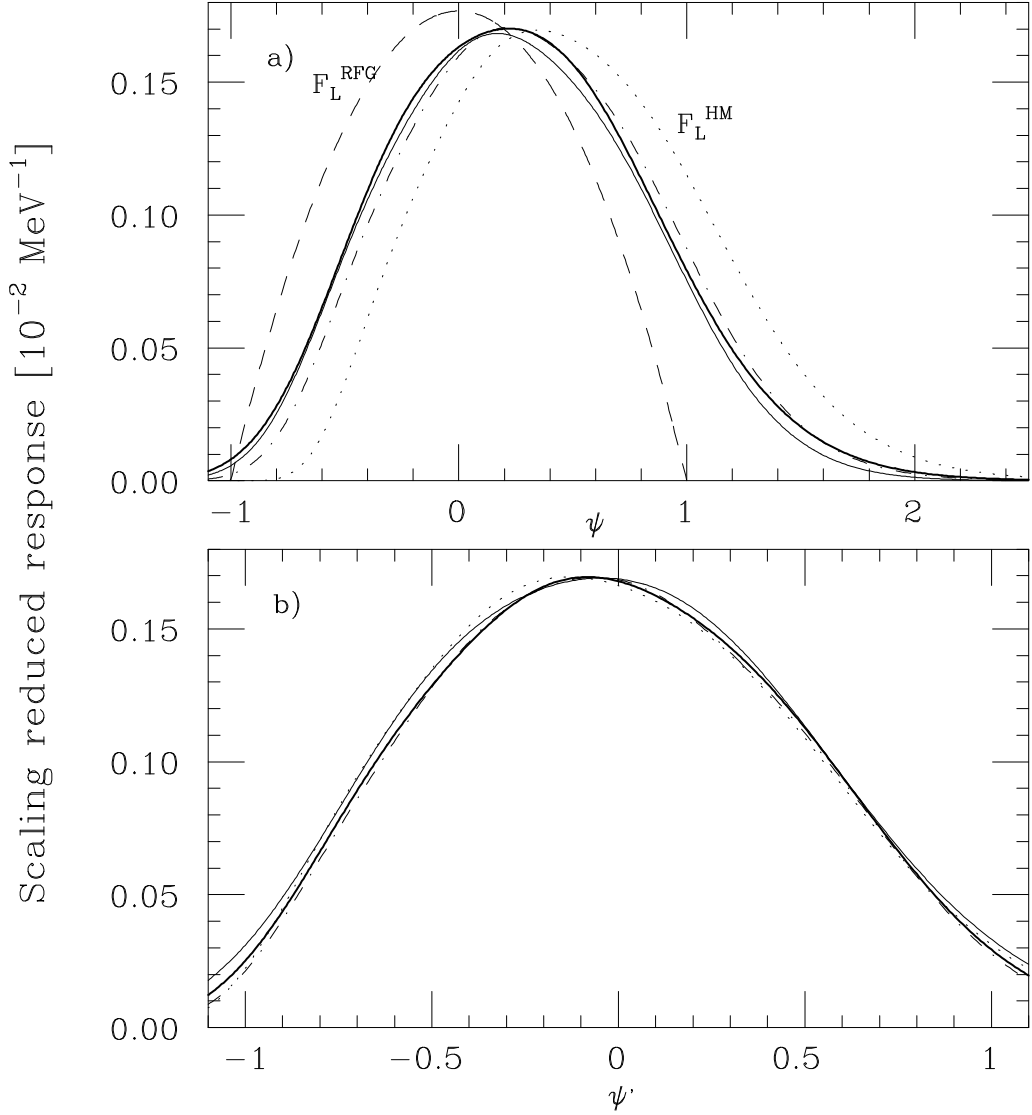,width=0.9\textwidth}}
\caption{The reduced response $F^{\text{HM}}_L$ (the hybrid model 
scaling function) is shown as a function of $\psi$ in panel (a) and 
$\psi'$ in panel (b) for 4 values of $q$ 
(fine dotted --- 0.5, dot-dashed --- 1.0, heavy solid --- 1.5 and
solid --- 2.0 GeV/c). The RFG result, which scales exactly as a function 
of $\psi$, is also shown for reference as a dashed curve in panel (a).
}
\label{fig:Fig4}
\end{center}
\end{figure}

\begin{figure}[tb]
\begin{center}
\mbox{\epsfig{file=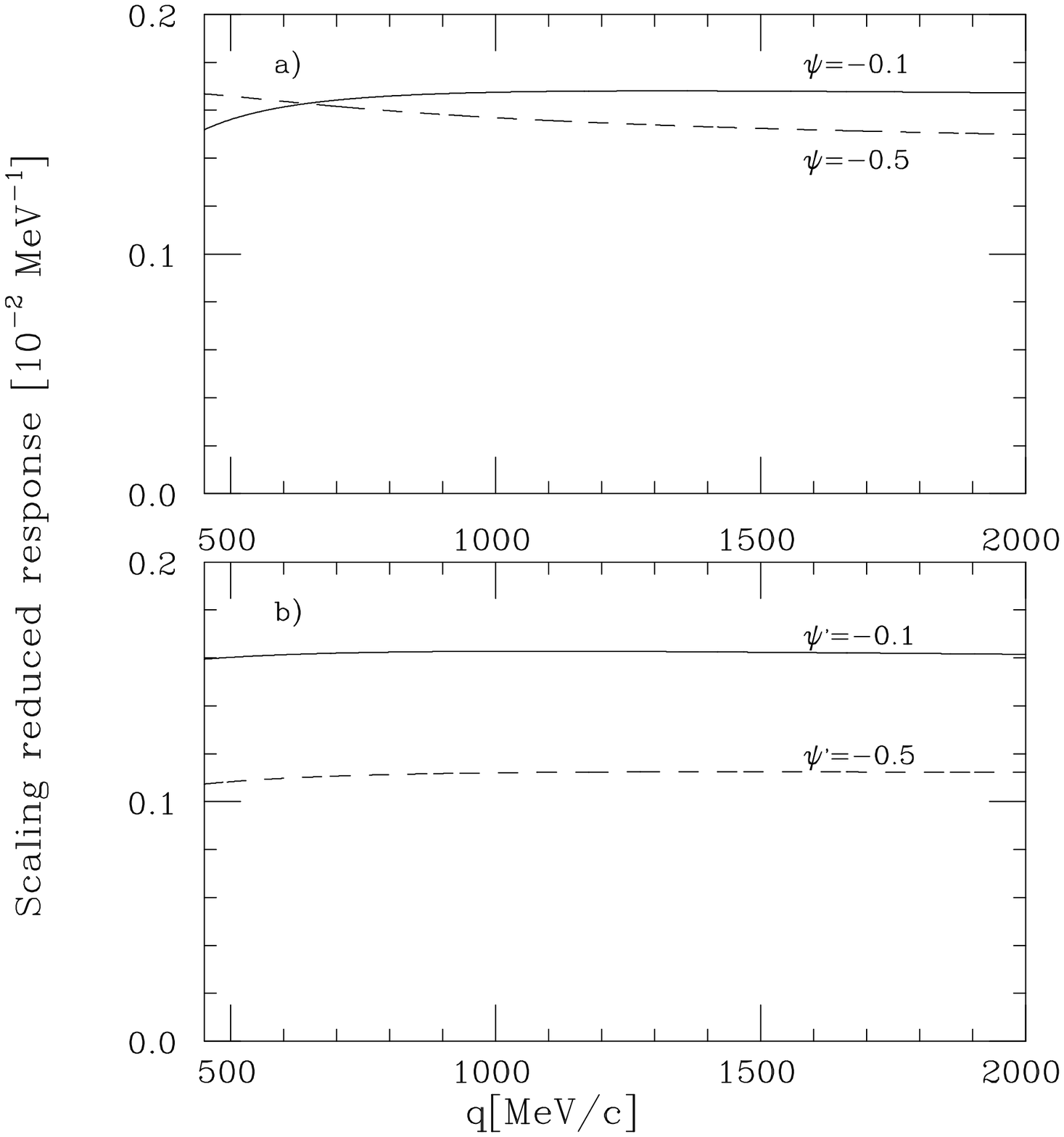,width=0.9\textwidth}}
\caption{The reduced response $F^{\text{HM}}_L$ is shown as a function
of $q$ at $\psi =-0.1$ and $-0.5$ in panel (a) and at $\psi' =-0.1$ and 
$-0.5$ in panel (b).
}
\label{fig:Fig5}
\end{center}
\end{figure}

The response function $R^{\text{HM}}_L$ of the hybrid model has been computed
in Ref.~\cite{Madre} and, when divided by $H_L$, shown to satisfy the CSR.
There, as already mentioned, it was seen that in setting up $\psi'$ the
effective Fermi momentum 237 MeV/c should be taken: the associated shift in 
$\lambda$ (see Eq.~(\ref{eq:shiftlam})) turns out then to be 
$\lambda_{\text{shift}}=$ 0.02. Fixing $G_L$ 
similarly, in Fig.~\ref{fig:Fig4} we display $F^{\text{HM}}_L$ for $^{40}$Ca 
versus $\psi$ and $\psi'$ for four different values of $q$ (the RFG result
is also shown for reference) and then display it in Fig.~\ref{fig:Fig5} 
versus $q$ for $\psi = -0.1$ and $-0.5$, as well as  for $\psi' = -0.1$ 
and $-0.5$. As seen in Fig.~\ref{fig:Fig4}, the HM scales either with $\psi$
or with $\psi'$ as $q$ becomes large; indeed, only the $q=$ 500 MeV/c plot 
versus $\psi$ shows any appreciable violation of scaling. Clearly
(by construction) the scaling versus $\psi'$ is excellent. It should also
be noted that the final scaling result in the upper panel in 
Fig.~\ref{fig:Fig4} lies a little to the right of $\psi=0$, while that in
the lower panel lies a little to the left of $\psi'=0$, suggesting that
the value for $\lambda_{\text{shift}}$ in Eq.~(\ref{eq:shiftlam}) used here
is somewhat too large and has led to an ``overshooting'' of the shift. 
Figure~~\ref{fig:Fig5} shows the evolution of scaling with $q$ and confirms
that the asymptotic behavior has essentially been reached by about
1 GeV/c. We also see that the scaling sets in sooner when $|\psi'|$ is
smaller, i.e., when one is closer to the quasielastic peak position.

\begin{figure}[tb]
\begin{center}
\mbox{\epsfig{file=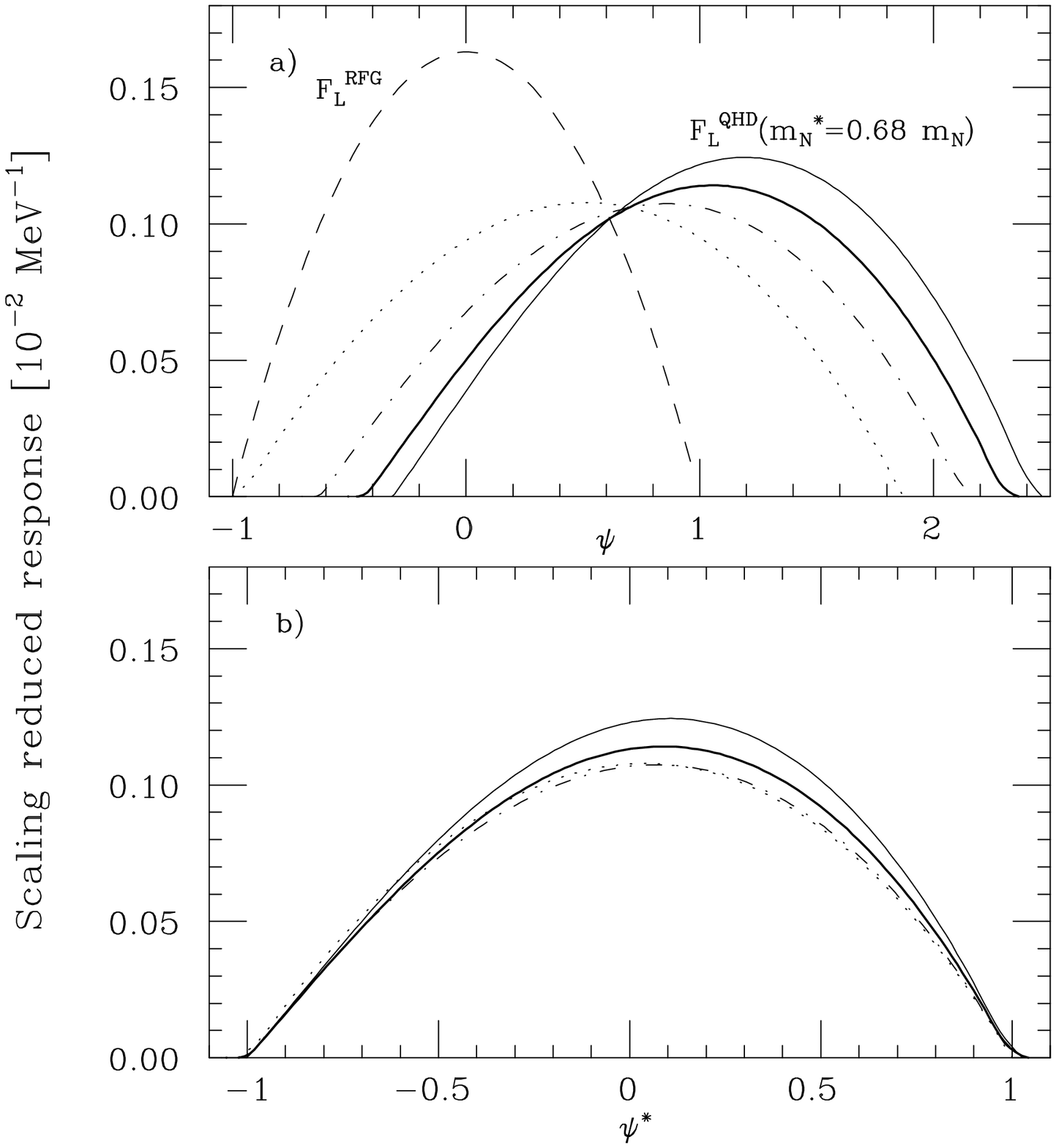,width=0.9\textwidth}}
\caption{The reduced response $F^{\text{QHD}}_L$ with $m_N^\ast =$ 0.68 $m_N$ 
is shown as a function of $\psi$ in panel (a) and 
$\psi^\ast$ in panel (b) for 4 values of $q$ 
(fine dotted --- 0.5, dot-dashed --- 1.0, heavy solid --- 1.5 and
solid --- 2.0 GeV/c). The RFG result, which scales exactly as a function 
of $\psi$, is also shown for reference as a dashed curve in panel (a).
}
\label{fig:Fig6}
\end{center}
\end{figure}

\begin{figure}[tb]
\begin{center}
\mbox{\epsfig{file=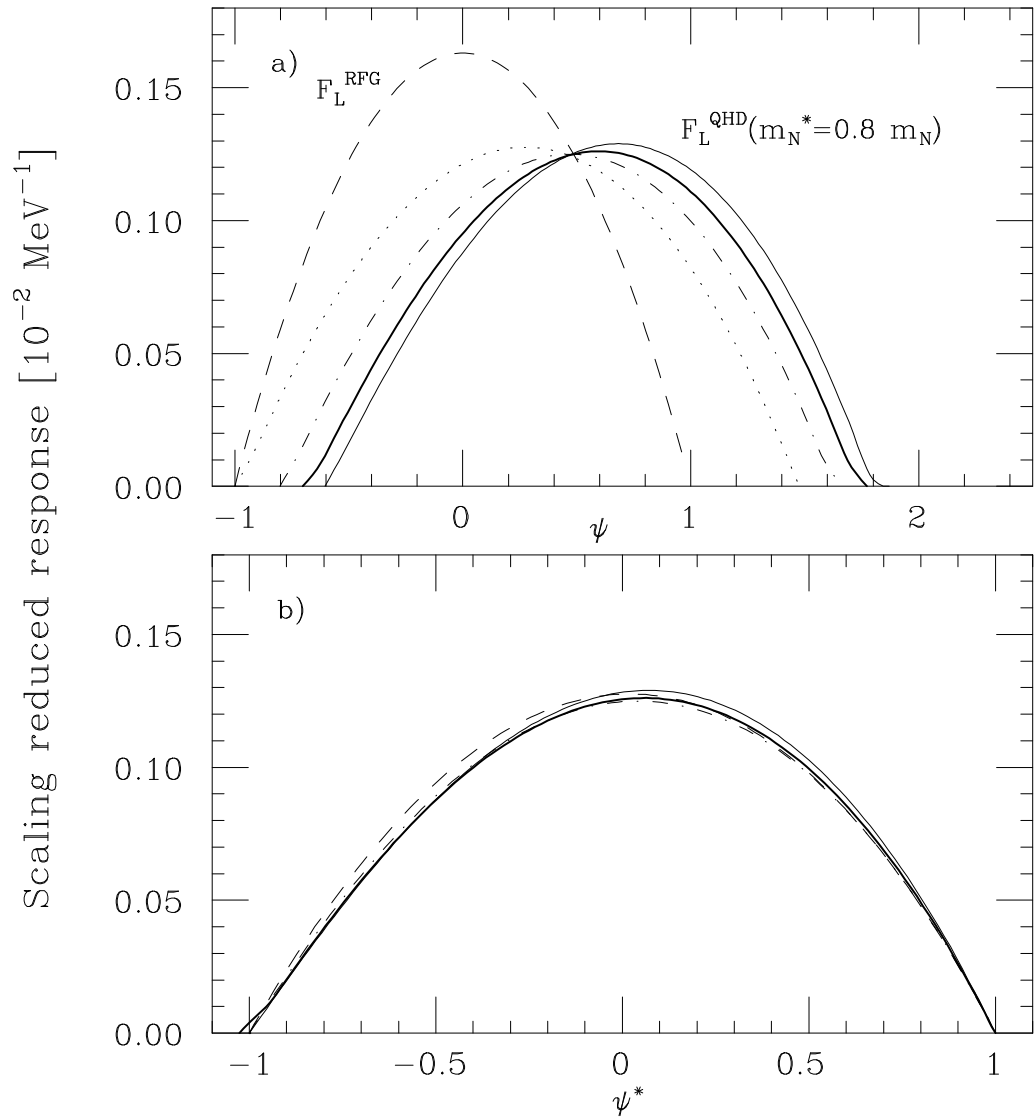,width=0.9\textwidth}}
\caption{As in Fig.~\ref{fig:Fig6}, except now with $m_N^\ast = 0.68\, m_N$
}
\label{fig:Fig7}
\end{center}
\end{figure}

\begin{figure}[tb]
\begin{center}
\mbox{\epsfig{file=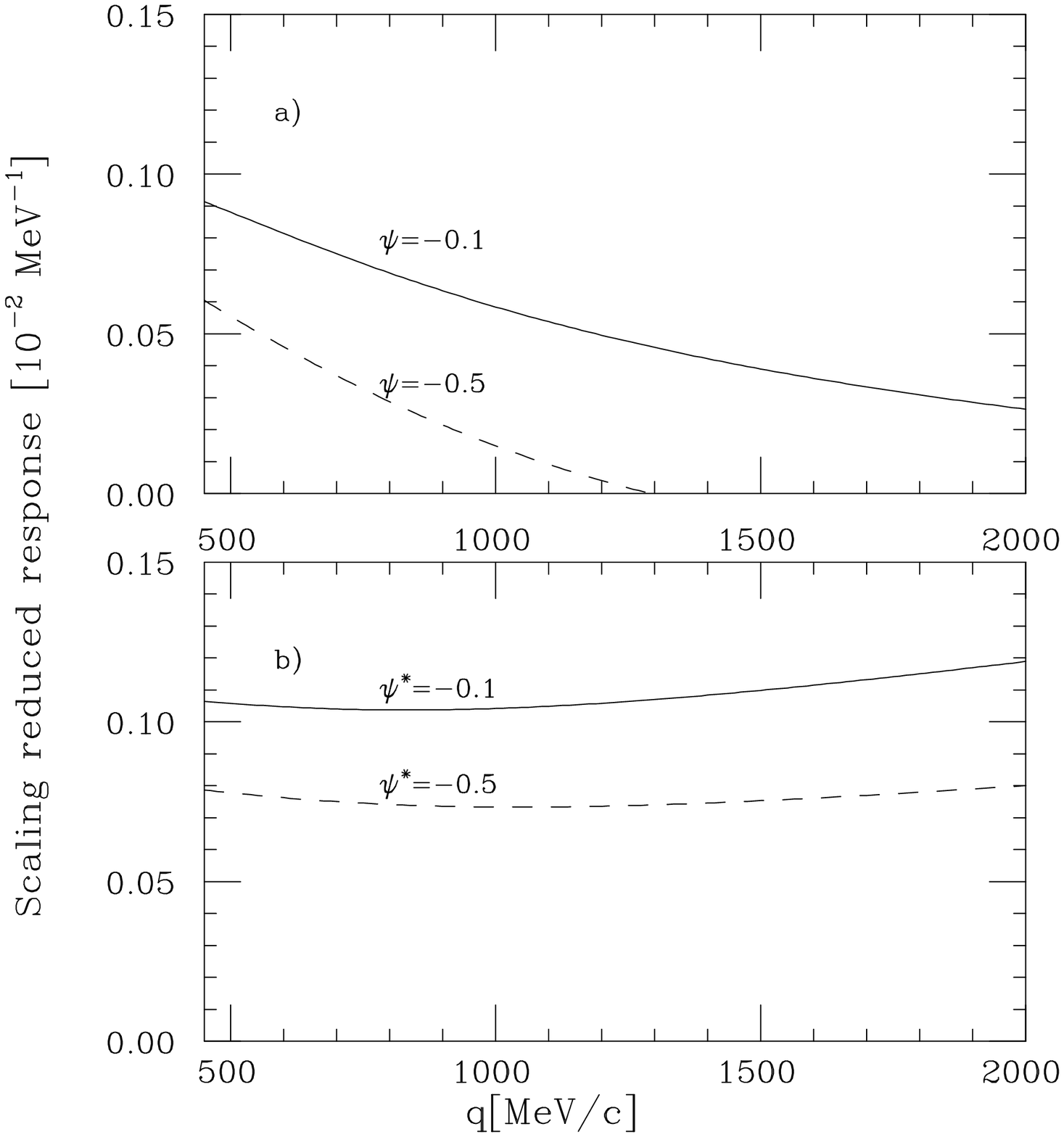,width=0.9\textwidth}}
\caption{The reduced response $F^{\text{QHD}}_L$ is shown as a function
of $q$ at $\psi =-0.1$ and $-0.5$ in panel (a) and at $\psi^\ast =-0.1$ and 
$-0.5$ in panel (b).
}
\label{fig:Fig8}
\end{center}
\end{figure}

In Figs.~\ref{fig:Fig6} and \ref{fig:Fig7} we display $F^{\text{QHD}}_L$ for 
$^{40}$Ca versus $\psi$ and $\psi^\ast$ for four different values of $q$ 
(the RFG result is also shown for reference) and then display it in 
Fig.~\ref{fig:Fig8} versus $q$
for $\psi = -0.1$ and $-0.5$, as well as  for $\psi^\ast = -0.1$ and $-0.5$.
Here results are shown both for $m_N^\ast = 0.68\,  m_N$ and 0.8 $m_N$.
The results in Figs.~\ref{fig:Fig6} and \ref{fig:Fig7}  show that 
$F^{\text{QHD}}_L$ does not scale versus $\psi$ when the effective mass
is constant and differs from $m_N$. As $q$ continues to grow beyond the range
of values shown in the figures, the results continue to shift to higher 
$\omega$ and never coalesce into a universal curve (see also 
Fig.~\ref{fig:Fig8}). When plotted versus $\psi^\ast$ the behavior, while
better, still does not scale. This is contrast with the RFG and HM results
displayed above and, importantly, is not what is seen experimentally where
the world data do appear to scale in $\psi$ \cite{TWDIS} (see also 
\cite{Will}). Comparing Figs.~\ref{fig:Fig6} and \ref{fig:Fig7}, it is
clear that scaling is better when $m_N^\ast$ is closer to $m_N$, indeed
becoming perfect when $m_N^\ast \to m_N$, as expected, since then 
$F^{\text{QHD}}_L \to F^{\text{RFG}}_L$. The fact that experimentally
the scaling is observed to occur successfully for $q$ greater than
about 1 GeV/c suggests that $m_N^\ast/m_N$ should not deviate appreciably
from unity for such kinematics.

\begin{figure}[tb]
\begin{center}
\mbox{\epsfig{file=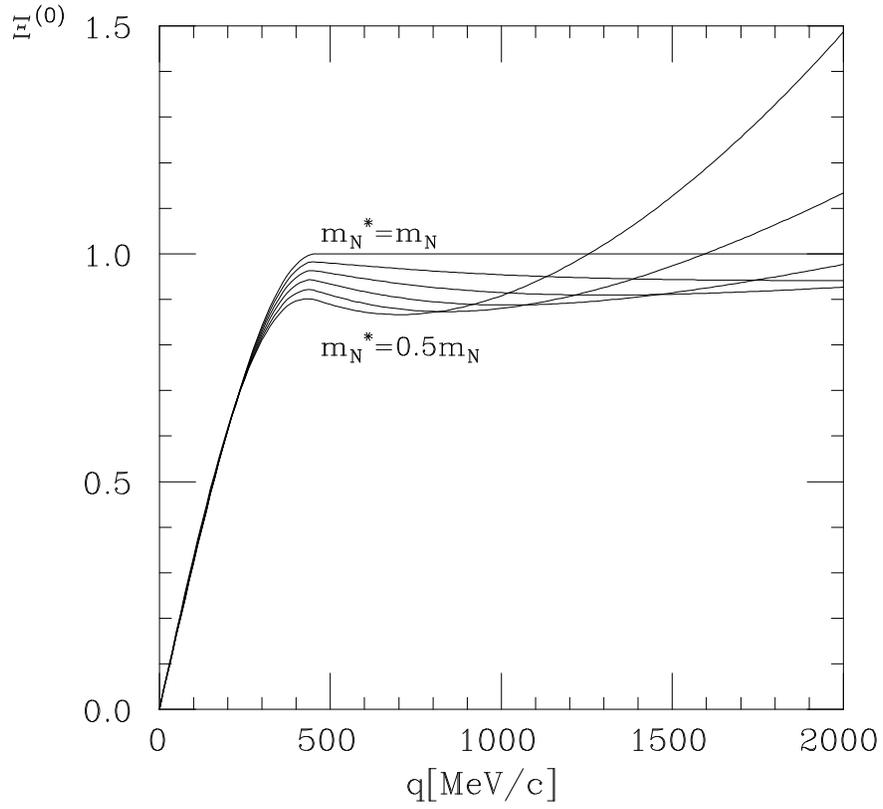,width=0.9\textwidth}}
\caption{Coulomb sum rule $\Xi^{(0),\text{QHD}}$ of the QHD model is shown 
as a function of $q$ for different values of the effective mass m$_N^\ast$, 
ranging from 1 to 0.5 $m_N$ in steps of 0.1 $m_N$. 
}
\label{fig:Fig9}
\end{center}
\end{figure}

In Fig.~\ref{fig:Fig9} the CSR associated with the QHD model is displayed
for  $^{40}$Ca ($k_F = 230$ MeV/c) and for a set of values of $m_N^\ast$
ranging from $0.5 m_N$ to the bare nucleon mass. Here we observe an 
effective mass dependence mostly due to the factor $m_N^\ast$
in front of the CSR integral 
\begin{eqnarray}
  \Xi^{(0),\text{QHD}} &=& \int\limits_0^q d\omega 
  r_L^{\text{QHD}} (q, \omega) \nonumber \\
  &=& \frac{3}{4} \left(\frac{m_N^\ast}{m_N}\right)^3
    \frac{\xi_F^\ast}{\xi_F} \int\limits_{-1}^1
    d\psi^\ast \frac{U_L^\ast}{U_L} (1 - \psi^{\ast^2})
    \frac{\partial \psi}{\partial\psi^\ast} \nonumber \\
  &\cong& \frac{3}{4} \frac{m_N^\ast}{m_N} \int\limits_{-1}^1
    d\psi^\ast \frac{U_L^\ast}{U_L} (1 - \psi^{\ast^2})
    \frac{\partial \psi}{\partial\psi^\ast} ,
\label{eq:Xi0star}
\end{eqnarray}
but also, to a less extent, to 
the Jacobian ${\partial\psi}/{\partial\psi^\ast}$. In fact, using 
Eq.~(\ref{eq:psiappr}), the latter can be cast in the form
\begin{equation}
  \frac{\partial\psi}{\partial\psi^\ast} \cong 1 
    - \frac{1 - m_N^\ast/m_N}{\sqrt{1 + 4 \kappa(\kappa + \eta_F \psi)}} ,
\label{eq:dpsidpsistarappr}
\end{equation}
which is less than unity for finite $\kappa$. These two
factors, which are both clearly related to many-body aspects of the 
nuclear response, are however partially counteracted by the 
``single-nucleon term'' $U_L^\ast/U_L$ which is also $m_N^\ast$
dependent and greater than unity. For large values of $\kappa$, 
Eq.~(\ref{eq:dpsidpsistarappr}) goes to one, whereas 
${U_L^\ast}/{U_L}$ continues to grow rapidly, increasingly so for smaller 
values of $m_N^\ast$. This then becomes the dominant factor in explaining 
the behavior of the curves in Fig.~\ref{fig:Fig9} for $q$, say, larger 
than 1 GeV/c. 

As discussed above, the expectation is that $m_N^\ast$ must evolve with
$q$ towards $m_N$. Thus, while at $q=$ 500 MeV/c a value of 
$m_N^\ast/m_N\cong$ 0.7 may be acceptable (implying that the 
resulting CSR will be about 5--10\% below unity), as $q$ reaches the
higher values between 1 and 2 GeV/c, values of  $m_N^\ast/m_N$ nearer
unity will be preferable from the point of view of scaling (and also
suggested by the full version of QHD, see Ref.~\cite{Horo}) and will
therefore yield a CSR that still remains close to unity.

\begin{figure}[tb]
\begin{center}
\mbox{\epsfig{file=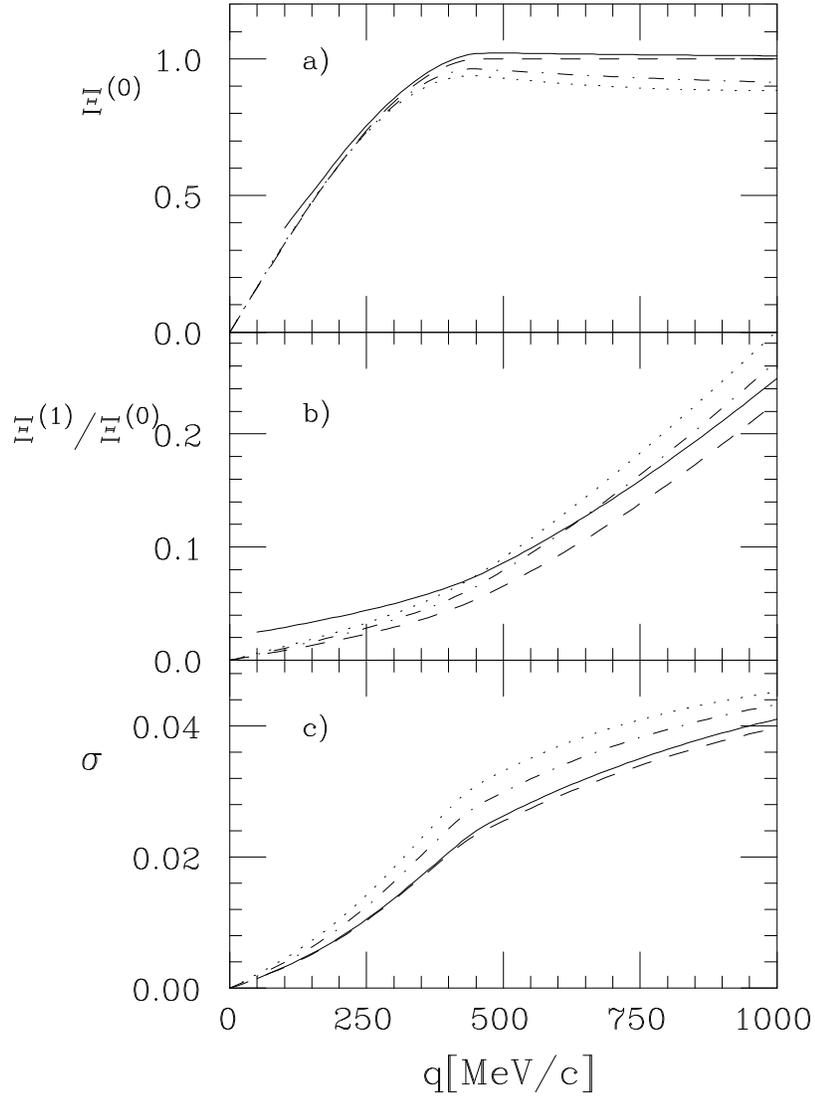,width=0.9\textwidth}}
\caption{The Coulomb sum rule (a), energy-weighted sum rule (b) and variance (c)
are shown as functions of $q$ for the three models employed in this work: 
dashed --- the RFG, solid --- the HM and the QHD model with 
$m_N^\ast$=0.8 $m_N$ (dot-dashed) and 0.68 $m_N$ (dotted). 
}
\label{fig:Fig10}
\end{center}
\end{figure}

For completeness, using the nomenclature introduced in Ref.~\cite{Madre}, we 
display in Fig.~\ref{fig:Fig10} the CSR ($\Xi^{(0)}$), the energy-weighted 
sum rule ($\Xi^{(1)}/\Xi^{(0)}$) and the variance 
$\sigma=\sqrt{\Xi^{(2)}-(\Xi^{(1)})^2}$ of $^{40}$Ca according to the HM, QHD
(with $m_N^\ast = 0.68$ $m_N$ and 0.8 $m_N$) and RFG models. As expected the 
responses of the HM and QHD models are considerably hardened with respect to 
that of the RFG; in addition the width of the QHD response is substantially 
wider than for the other two models. 

\begin{figure}[tb]
\begin{center}
\mbox{\epsfig{file=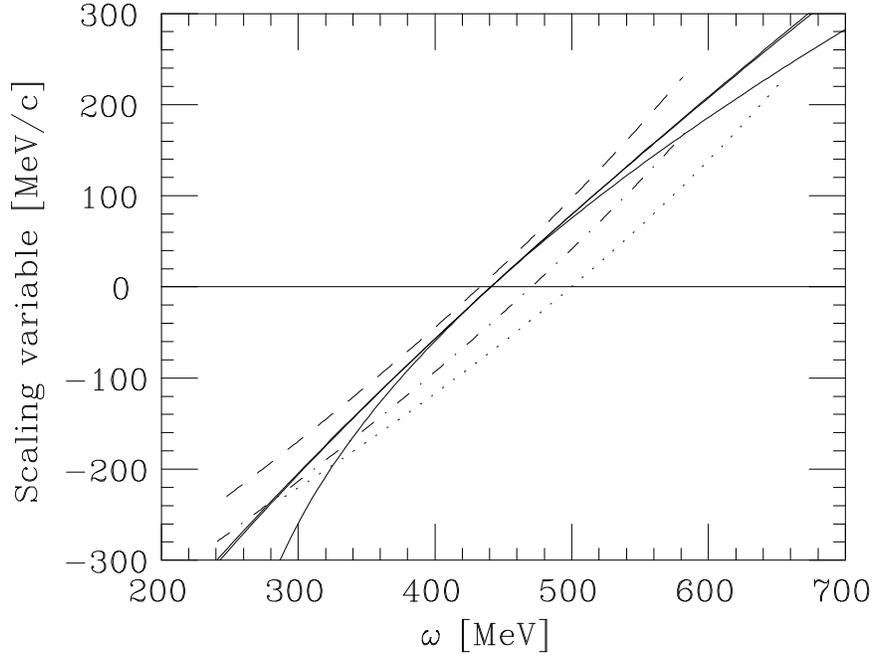,width=0.9\textwidth}}
\caption{The various scaling variables employed in this work are 
shown as functions of $\omega$ at $q=1$ GeV/c: $k_F\psi$ (dashed), 
$k_F (HM) \psi^\prime$ with $k_F (HM)=$ 237 MeV/c 
(dot-dashed), $k_F\psi^\ast$ (dotted) and $y$ (solid). 
The three solid curves correspond to $A=2$, 20 and 200,
respectively, starting from below. 
}
\label{fig:Fig11}
\end{center}
\end{figure}

In concluding this section we shortly discuss the scaling variables.
In this paper four scaling variables have been introduced: the canonical
$y$, the $\psi$ of the RFG (together with its related variations, $\zeta$ and
$y_{\text{RFG}}$), the $\psi^\prime$ of the HM and the $\psi^\ast$
of the QHD. How are they interrelated? To answer this question 
in Fig.~\ref{fig:Fig11} we display their behavior as functions of $\omega$ 
for $q=1$ GeV/c. The following features emerge from the figure:

a) The behavior of the canonical $y$ variable versus the mass number $A$ 
   is very rapid at small $A$ and appears by $A$ = 20 to have almost reached
   its asymptotic $A = \infty$ value. 

b) In accord with the findings of \cite{Madre}, for large
   $A$, $y$ is roughly given by Eq.~(\ref{eq:psiappr}) with $\lambda$ shifted 
   downward by an amount {\em grosso modo} corresponding to $E_S$. Hence in 
   the $A=\infty$ limit one does not recover the RFG $y$-variable from $y$,
   but a slightly shifted result.

c) The variable $\psi^\prime$ is also shifted downward with respect to $\psi$, 
   by an amount set by $\lambda_{\text{shift}}$ in Eq.~(\ref{eq:shiftlam}), 
   although we conjecture (see above) that this shift is somewhat too large.

d) Finally, note the behavior of $\psi^\ast$ with $\omega$: it
   turns out to be shifted towards higher energies with respect to the other
   scaling variables. This goes in parallel with the energy-weighted sum
   rule of the QHD model whose reduced response is substantially hardened 
   in comparison with other models here considered for $q\geq$500 MeV/c.

\section{ Final comments}
\label{sec:comm}

Our motivations in this work have been to study the Coulomb sum rule and
scaling in a unified way, and to explore various models for the inclusive
electromagnetic response of nuclei in the quasielastic region to see
whether or not they simultaneously satisfy the CSR and scale. We have
demonstrated how our previous work on the CSR may be generalized to 
include treatments of scaling: in both cases the response functions or
the cross section are divided by specific functions, $H_L$ for the CSR and 
$G_L$, $G_T$ or $\sigma_M (v_L G_L + v_T G_T)$, as is appropriate, for
scaling, to yield reduced responses with the desired properties, namely,
a CSR or scaling behavior. We have studied the model-dependence of these
dividing functions within the context of two specific models for the
single-nucleon content in the problem and found it to be small, as long as
we restrict our attention to high momentum transfers and to the vicinity
of the quasielastic peak. Essentially, for such conditions, 
we have obtained nearly universal dividing functions with which any model
can be tested and with which experimental data can be reduced. Typically
we find corrections at high $q$ which are characterized by $\eta_F^2=
(k_F/m_N)^2$, namely, reaching only $\sim$8\% for the heaviest nuclei.

In the course of this study we have introduced and inter-related several 
types of scaling variables ($y$, $\psi$, $\zeta$, $y_{\text{RFG}}$, $\psi'$).
These variables are all closely related: they differ at most by small shifts 
introduced by the various models to account (at some level) for interaction 
effects. On the one hand, when experimental data are reduced using our 
dividing functions, they are seen to yield a CSR at high-$q$ and to scale 
when plotted versus any of the scaling variables listed above. On the other
hand, the models considered in this work yield various results. The
relativistic Fermi gas (RFG) model and the hybrid model (HM) both saturate
the CSR at high-$q$. In contrast, the quantum hadrodynamical (QHD) model
does so only if the effective value of $m_N^\ast/m_N$ evolves with 
increasing $q$ towards unity, as suggested by the latest version of the
model. With regard to scaling, the RFG model (by construction) scales with 
$\psi$ (and of course with $\zeta$ or $y_{\text{RFG}}$, since they are 
intrinsically related). It also scales quite well versus $y$. The HM scales 
very well with $\psi'$ (by construction), but also quite well with $\psi$.
In contrast, the QHD model does not scale with any of the above variables
if the ratio $m_N^\ast/m_N$ is constant and differs significantly from
unity. One must conclude that the successful scaling behavior seen 
experimentally indicates that $m_N^\ast$ must approach $m_N$ as $q$
increases beyond about 1 GeV/c.

Finally, we note that an exact CSR and exact scaling behavior should not
be expected to occur. We certainly believe that interaction effects beyond
those of the mean field (whose effects are incorporated at least to some extent 
in the models studied here) can play a role even at momentum 
transfers as high as 1 GeV/c. Furthermore, we expect that two-body meson
exchange current effects can also play a role, for instance, making the
longitudinal and transverse reduced responses scale to different functions.
Both of these classes of corrections are currently being re-investigated.

\acknowledgements{
This work was supported in part by funds provided by the U.S. Department
of Energy (D.O.E.) under cooperative research agreement
\#DF-FC02-94ER40818 and by the INFN-MIT ``Bruno Rossi'' Exchange Program.}


\newpage

\newpage

\newpage

\newpage

\newpage

\newpage

\newpage

\newpage

\newpage

\newpage

\newpage

\end{document}